\documentclass{elsart}

\usepackage{harvard}

\usepackage{graphicx}

\usepackage{amssymb}

\def\url#1{{\ttfamily\def\/{/\discretionary{}{}{}}#1}}

\begin{document}


\begin{frontmatter}
  \title{Rich-Cluster and Non-Cluster Radio Galaxies \& The (P,D)
    Diagram for a Large Number of FR I and FR II Sources} 
  
  \author[Ledlow]{M.J. Ledlow\thanksref{ml}}, \author[Owen]{F.N.
    Owen}, \author[Eilek]{J.A. Eilek}
  
  \thanks[ml]{E-mail: mledlow@wombat.phys.unm.edu}
  
  \address[Ledlow]{Institute for Astrophysics, Dept. of Physics \&
    Astronomy, University of New Mexico, 800 Yale Blvd. NE,
    Albuquerque, NM 87131}
  
  \address[Owen]{National Radio Astronomy Observatory\thanksref{nrao},
    P.O.  Box 0, 1003 Lopezville Road, Socorro, NM 87801}
  
  \address[Eilek]{Astrophysics Research Center, Dept. of Physics, New
    Mexico Institute of Mining \& Technology, Socorro, NM 87801}
  
  \thanks[nrao]{The National Radio Astronomy Observatory is operated
    by Associated Universities, Inc., under contract with the National
    Science Foundation.}

\begin{abstract}
  We present a comparison of the optical and radio properties of radio
  sources inside and outside the cores of rich clusters from combined
  samples of more than 380 radio sources.  We also examine the nature
  of FR I and FR II host galaxies, and in particular, we illustrate
  the importance of selection effects in propagating the misconception
  that FR I's and FR II's are found in hosts of very different optical
  luminosity.  Given the large sample size, we also discuss the
  power-size ($P,D$) distributions as a function of optical
  luminosity.
\end{abstract}

\end{frontmatter}

\section{Introduction}

The primary goal of this work is to compare radio sources in and out
of the rich cluster environment to look for differences in either the
host galaxies or the distribution of sources in radio luminosity,
size, or morphology.  The surprising result that the amplitude and
shape of the radio luminosity function (RLF) is identical inside and
outside the cores of rich clusters \cite{ledl96,fanti84}, is puzzling
given what we know about the X-ray properties of galaxy clusters and
groups.  While poor clusters may simply be a lower-mass extension of
rich clusters, given the differences in both ambient external density
and local velocity dispersion, one might expect a significantly
different rate of galaxy interactions, a higher frequency of gas-rich
neighbors, and a lower density medium to confine the radio sources.
The similarity of the RLF in and out of clusters suggests that the
same frequency of elliptical galaxies are radio emitting (also with a
similar distribution in radio powers).  It does not, however, rule out
a different relation in either the radio power-size ($P,D$) plane or
the relationship between radio and optical luminosity and the FR I/II
division \cite{deruiter90,ledl95b}.

In order to study these effects, we have put together the largest
available samples of radio sources to date, including 259 rich-cluster
radio galaxies \cite{ledl95b,ledl96} and 124 radio sources not found
within 1 Abell radius ($\sim 2h_{75}^{-1}$ Mpc) of an Abell cluster.
The non-cluster samples are based on the Bologna B2 Survey
\cite{fanti87} and a subset of the all-sky 2.7GHz survey from
\citeasnoun{wp}(WP).  We include here analysis of new optical (R-band)
observations of all 124 non-cluster radio sources, which were analyzed
identically to our rich cluster survey \cite{ledl95b}.  We have also
used the NRAO VLA D-Array Sky-Survey (NVSS) to measure source fluxes
(and in some cases sizes) for the B2 and WP samples.  Radio sizes for
the rich-cluster radio galaxies are from \citeasnoun{eilek}.

\section{Selection Effects}
\label{selection}

Before we can compare our two samples (cluster \& non-cluster), we
need to explore the effects of the selection function on the various
samples.  The B2 sample has traditionally been divided into B2-Near
\cite{b2n} and B2-Far \cite{b2f} subsamples, each with a different
selection function.  The near-sample has a magnitude-limit of $m_{pg}<
15.7$ ($m_{24.5}^R \lesssim 14$).  The far-sample has a magnitude
cutoff of $m_v=16.5$ ($m_{24.5}^R \lesssim 15.9$).  Both subsamples
essentially have the same radio flux-limit of $P_{408MHz}=200$ mJy
($\approx 100$ mJy assuming $F_{\nu} \propto \nu^{-0.7}$).  The WP
sample has no magnitude cutoff, but has a high flux-limit of
$P_{2.7GHz}=2$ Jy ($P_{1400MHz} \geq 3.2$ Jy, also for $\alpha=0.7$).

\begin{figure}
\begin{center}
  \includegraphics*[width=10cm]{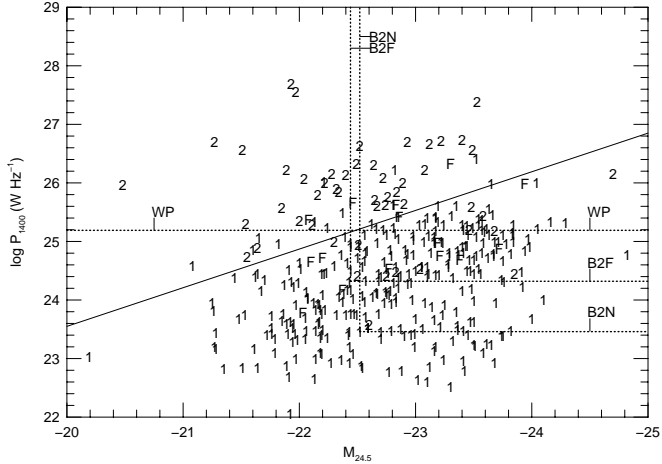}
\end{center}
\caption{FR I/II diagram in the optical/radio luminosity plane. The dotted
  lines show the 75th percentile values of $M_{24.5}$ and $P_{20cm}$
  for the various samples. The symbols are: FR I=1's, FR II=2's, and
  Fat-Double=F. The solid-line indicates the fiducial FR I/II break.}
\label{selectioneffects}
\end{figure}

The effect of these selection functions is illustrated in Fig.
\ref{selectioneffects} where we have shown the 75th percentile values
of $M_{24.5}^R$ and $P_{1400}$ for each of the samples (75\% of the
objects are to the right of the vertical lines and above the
horizontal lines).  Because of the magnitude-limits of the B2 sample,
the part of the optical/radio luminosity plane fainter than $M^{*}_R =
-22$ and $P_{1400}<10^{24}$ is essentially unsampled.  This net effect
of these selection functions is to preferentially pick out FR I radio
sources with optical luminosities $\gg L^*$.  {\bf It is for these
  reasons that FR I's have mistakingly been identified with only the
  brighter ellipticals, consistent with brightest cluster galaxies
  (BCG)}.  It is clear from Fig. \ref{selectioneffects} that FR I's
actually span a range of more than a factor of 40 in optical
luminosity, and trace the normal optical luminosity function for
elliptical galaxies \cite{ledl95b}.  Additionally, these same
selections explain why FR II's have always been associated with $M^*$
ellipticals.  Given the steep slope of both the optical and radio
luminosity functions for $M<-23$ and $P_{1400}>10^{25}$ and the small
search volumes, all detected FR II's are likely to cluster near the
survey-limits in optical luminosity (near $M^*$). It should be noted,
however, that it is difficult to separate out the effects of redshift
evolution with a limited search volume.

\section{Rich Cluster vs. Non Rich-Cluster Radio Galaxies}

In order to compare our rich cluster radio sources to these other
samples, we have therefore applied the B2 selection function (both
$M_{opt}$ and $P_{20cm}$ constraints) to our rich cluster survey,
including an upper-redshift cutoff of z=0.24 to better match the two
samples.  We have compared a number of parameters from both the
optical analysis as well as the radio properties between these two
samples.  See \citeasnoun{ledl2000} for a complete presentation of
these results.

For the optical properties, we compared the distribution of
ellipticities, surface-brightness profile shape (via a power-law
exponent and $r^{1/4}$-law fits), departure from elliptical isophotes
(A4,B4 parameters), the Kormendy relation ($\mu_e$ vs. $r_e$ slope),
and the luminosity-size ($M_{24.5}$ vs.  $r_{24.5}$) relationships for
both the cluster and non-cluster samples.  Measured at the same
optical luminosity ($M_{24.5}$), we find no discernable differences
between the two samples.

We also compared the optical properties of the FR I and FR II host
galaxies (combining cluster and non-cluster sources together). From
the median of the optical magnitude distributions, we find:
$M_{24.5}(FR I)=-22.85\pm0.04$ and $M_{24.5}(FR II)=-22.58 \pm0.10$.
Thus, the FR I's are found in only slightly brighter galaxies, and
significantly less luminous on average than BCG's ($M_{24.5} <
-23.5$).  Additionally, we found no differences in the host galaxy
properties listed above between the FR I and FR II hosts, at least in
their broadband photometric properties.

Our radio comparison between cluster and non-cluster sources included
the size-distributions (at the same radio power), the distribution in
the radio/optical luminosity plane, and the FR I/II division.  We find
no evidence for differences in either of these properties between the
two samples. We do however note the following observations: 1) The
size-distribution is somewhat broader outside of clusters, and 2) We
do not see a population of FR II's with $P_{20cm} \gtrsim 25.5$ in the
rich cluster sample. However, these differences are minor, and argue
for very similar envirnoments (at least the aspect of the environment
which most influences the evolution of radio sources) inside/outside
the cores of rich clusters.  All three samples are plotted together in
Fig.\ref{selectioneffects} and Fig.  \ref{pdplots}.

\section{The (P,D) Diagram} 

\begin{figure}[h]
\begin{center}
\includegraphics*[height=12cm]{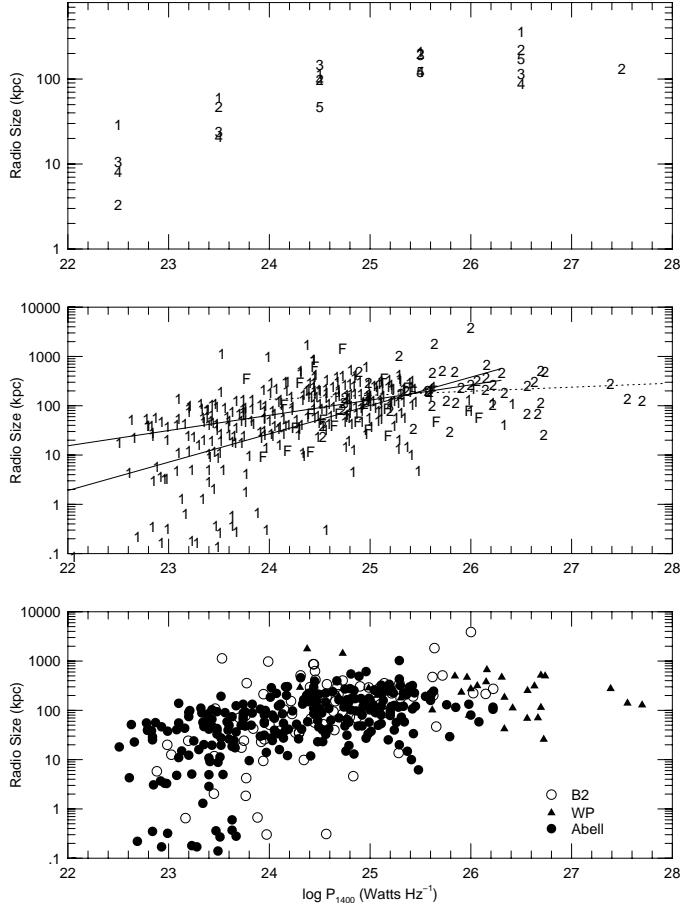}
\end{center}
\caption{(middle) ($P,D$) diagram for combined cluster/non-cluster samples with a 
  cutoff of z=0.24.  The solid-lines show the fit for the FR I's with
  and without the size upper- limits for the unresolved sources.  The
  dashed-line is the fit for the FR II's. (bottom) Same as middle,
  where the point types reflect the sample from which the sources were
  taken (see legend on plot). (top) Median values of the radio size in
  radio power bins as a function of optical magnitude ($\Delta M$=0.75
  mag); centered on 1=-21.38, 2=-22.13, 3=-22.88, 4=-23.63, 5=-24.38.}
\label{pdplots}
\end{figure}

Given that we find no significant differences in our
cluster/non-cluster samples, we have combined them in order to improve
the statistics.  In Fig. \ref{pdplots}, we show the radio-power/size
($P,D$) distributions in several ways.  In the bottom diagram we
separate the three samples by point-type to show their intrinsic
distributions.  In the middle plot we show the sample divided into FR
I, FR II, and Fat-double classes.  For the FR I's, we find a fit
(log-log) of the form: $Size (FR I) \propto P_{20cm}^{0.31\pm0.03}$
(ignoring unresolved sources with only size upper-limits).  For the FR
II's, we find essentially no dependence between power and size ($Size
(FR II) \propto P_{20cm}^{0.08\pm0.09}$).  Note, however, that at the
same radio power, both FR I and FR II sources have the same
distribution in source-sizes.

The top plot in Fig. \ref{pdplots} shows the median size as a function
of radio power and optical luminosity (lower numbers represent bins of
lower optical luminosity).  One sees from the plot that the trend is
for optically fainter galaxies to have larger radio sources. In bins
of constant optical magnitude, we find that for fits of the form:
$Size \propto P_{20cm}^x$, x decreases with increasing $L_{opt}$ (from
0.36-0.29).  We also find that the size distributions for $M\gtrsim
M^*$ and $M\lesssim -23$ are different at the 98\% level.  It thus
appears that optically brighter galaxies tend to have smaller sizes
when measured at the same radio power.  Is this result a consequence
of environment (the brighter galaxies being in a higher-density
environment or near the bottom of the local gravitational potential)
or initial conditions (the physics near the black hole and the jet
initial conditions)?  We plan to address these issues by comparing our
observed distributions in ($M_{opt},P_{20cm},Size$) with model
predictions.  It is clear, however, that we may need a
new/more-detailed model for FR I sources (see \citeasnoun{eileklc}
these proceedings) in order to understand these results.

\end{document}